\documentclass[aps,prl,twocolumn,showpacs,superscriptaddress,groupedaddress]{revtex4}

\usepackage{bbm}

\usepackage{graphicx}
\usepackage{dcolumn}
\usepackage{amsmath}
\usepackage{bm}
\usepackage{color}
\usepackage{mathrsfs}
\usepackage{epstopdf}
\usepackage{amsmath}
\usepackage{amssymb}
\usepackage{subfigure}
\usepackage{enumitem}
\usepackage{multirow}
\usepackage{exscale}
\usepackage{relsize}
\usepackage{dsfont}

\newcommand{\be}{\begin{equation}}
\newcommand{\ee}{\end{equation}}
\newcommand{\bey}{\begin{eqnarray}}
\newcommand{\eey}{\end{eqnarray}}
\newcommand{\bw}{\begin{widetext}}
\newcommand{\ew}{\end{widetext}}

\newcommand{\ba}{\begin{array}}
\newcommand{\ea}{\end{array}}
\newcommand{\bi}{\begin{itemize}}
\newcommand{\ei}{\end{itemize}}
\newcommand{\bem}{\begin{enumerate}}
\newcommand{\eem}{\end{enumerate}}

\begin{document}

\title{Quantum stochastic thermodynamics: A semiclassical theory in phase space}

\author{Zhaoyu Fei} \email[Email address: ]{1501110183@pku.edu.cn}
\affiliation{Department of Physics and Key Laboratory of Optical Field Manipulation of Zhejiang Province, Zhejiang Sci-Tech University, Hangzhou 310018, China}
\affiliation{Graduate School of China Academy of Engineering Physics,
No. 10 Xibeiwang East Road, Haidian District, Beijing, 100193, China}

 \date{\today}

\begin{abstract}
A formalism for quantum many-body systems is proposed through a semiclassical treatment in phase space, allowing us to establish stochastic thermodynamics incorporating quantum statistics.
%By semiclassically treating quantum many-body systems, a stochastic thermodynamics involving quantum statistics is established.
%We consider a stochastic  Fokker-Planck equation as the dynamics in  mesoscopic level.
Specifically, we utilize a stochastic Fokker-Planck equation as the dynamics at the  mesoscopic level. Here,
the  noise term characterizing the fluctuation of the flux density  accounts for the  finite-$N$ effects arising from random collisions between the system and the reservoir. Accordingly, the stationary solution is a quasi-equilibrium state in a canonical system. %instead of the equilibrium state.  %consistent with the theory of equilibrium fluctuation.
%We firstly demonstrate the corresponding microscopic dynamics %, a system of $N$-particle Ito-Langevin equations,
%by assuming the correlation between thermal noises of different particles.
We define  stochastic thermodynamic quantities based on the trajectories of the phase-space distribution. %, rather than trajectories  of particles in previous studies. %, which is beyond  the previous scenarios of stochastic thermodynamics based on trajectories of particles.. %and the corresponding $N$-particle microscopic interpretation. %, which reproduces their counterparts for a one-particle system.
The  conservation law of energy, $H$ theorem and  fluctuation theorems are therefore obtained. %And, the connection between stochastic thermodynamics and the quasi-equilibrium state is established.
Our work sets an alternative formalism of quantum stochastic thermodynamics that is independent of the  two-point measurement scheme. The numerous projective measurements of  quantum systems are replaced by the sampling of the phase-space distribution,  offering hope for experimental verifications in the future. %when the appropriate techniques become available.

\end{abstract}

\maketitle

\textit{Introduction}.---Stochastic thermodynamics uses stochastic variables to better understand the non-equilibrium dynamics present in microscopic systems.
%Stochastic thermodynamics aims to extend thermodynamics to systems that operate far from equilibrium and are subject to thermal fluctuations.
In these systems, %(such as molecular motors, Brownian particles, and chemical reaction networks),
thermal fluctuations are significant and the laws of thermodynamics need to be understood from probabilistic perspectives. Typical results include the definition of stochastic thermodynamic quantities and fluctuation theorems~\cite{Sekimoto2010, Seifert2012, Klages2013, Crooks1999}, the latter of which quantify the statistical behavior of nonequilibrium systems  and generalize the second law. In addition,  exact thermodynamic statements beyond the realm of linear response are obtained. These exact results
refer to distribution functions of thermodynamic quantities such as exchanged heat, applied work, and entropy production for these systems~\cite{Seifert2012}.

Stochastic thermodynamics is also applied to quantum systems by usually introducing the two-point measurement scheme~\cite{aq2000}. Here, work is not an observable~\cite{Talkner1999}, but is defined as the energy difference between the initial and final projective measurements. By applying the two-point measurement scheme to open quantum systems, stochastic thermodynamics is also established for Lindblad-type systems~\cite{Hekking2013} and Caldeira-Leggett-type systems~\cite{Funo2018}. For a comprehensive overview of quantum stochastic thermodynamics, readers are referred to Ref.~\cite{Strasberg2021}. %is a reference of knowing quantum stochastic thermodynamics.

%, which leads to both the conservation of energy and the fluctuation theorems (see reviews of quantum stochastic thermodynamics in Ref.~\cite{Strasberg2021}).
%%We refer Ref.~\cite{Strasberg2021} as a good book about the (classical) quantum stochastic thermodynamics.
%Besides, the path-integral approach to the two-point measurement scheme is given in Refs.~\cite{Funo2018}.
%However, such a scheme has a fatal weakness that it is unpractical to make numerous projection measurements on a  quantum many-body system or the  reservoir in a realistic experiment. Hence, it is necessary to establish a quantum stochastic thermodynamics within other formalisms.
The two-point measurement scheme assumes numerous projective measurements of quantum systems. However, such an assumption is impractical for %the impracticality of numerous projection measurements on
quantum many-body systems or the  reservoir in realistic experiments, presenting a significant challenge to this scheme. Therefore, it is necessary to establish quantum stochastic thermodynamics within alternative formalisms.

In our view, the reason for the conundrum is that %the projective measurements are so precise that
too much information is involved in projective measurements to establish quantum stochastic thermodynamics.
We circumvent this issue by semiclassically treating  quantum systems in   phase space. When the
largest energy-level spacing of the system is small compared to the thermal excitation energy,  a phase-space description of the system under a proper error is possible through a distribution function $\rho(z),z=(x,p)$ (a one-dimensional system is considered here for simplicity)~\cite{foot1}. %The position and its canonical momentum commutes under a proper error.
Meanwhile, the temperature is low enough that the system still presents  quantum statistics,
% the Bose-Einstein statistics for bosons and Fermi-Dirac statistics for fermions,
i. e., the thermal wavelength of a particle is comparable to the interparticle spacing~\cite{Pathria2011}.

In our formalism, the system is governed by a stochastic  Fokker-Planck equation that is a nonlinear %Fokker-Planck
equation incorporating a noise term that characterizes the fluctuation of the flux density. The nonlinear equation %in contrast to its linear form that implies the Maxwell-Boltzmann statistics,
determines an equilibrium state satisfying   non-Boltzmann statistics, especially for  quantum systems~\cite{Kaniadakis2001, Frank2005, Chavanisa2008}. A simple derivation using heuristic arguments is provided in Ref.~\cite{Kadanoff2000}. As an approximation of quantum Boltzmann equations, such a nonlinear equation has been used to study finite fermionic or bosonic systems~\cite{Wolschin1982, Wolschin2018},  and the electron collisions in dense plasmas~\cite{Daligault2016} . The  noise term %, characterizing the fluctuation of the particle-number flux,
keeps track of the finite-$N$ effect arising from random collisions between the system and the reservoir. Consequently, the system implies a quasi-equilibrium state in a canonical system, %maintains that the steady state of the  probability distribution of observing the phase-space distribution $\rho_t(z)$ is quasi-equilibrium  in a canonical system~\cite{Mishin2015}.
which originates from the reverse form   of the Boltzmann entropy, i. e., the exponential of the entropy denotes the number of the corresponding microscopic states according to Einstein's interpretation~\cite{Mishin2015, Einstein1910}.

In this paper, we establish the stochastic thermodynamics based on the trajectories of the phase-space distribution. %,  rather than trajectories  of particles in previous studies. %which is beyond  the previous scenarios of stochastic thermodynamics based on trajectories of particles. %whose probability distribution is determined by a functional Fokker-Planck equation.
Accordingly, the stochastic thermodynamic quantities  are defined %on the trajectory of phase-space distribution
and reproduce their counterparts in previous studies for systems consisting of distinguishable noninteracting particles. The conservation law of energy, the $H$ theorem, and  fluctuation theorems %(detailed form and integral form)
are therefore obtained. %according to these stochastic thermodynamic quantities.
A connection between stochastic thermodynamics and the quasi-equilibrium state is also established.

For  experimental verifications of our formalism, we would like to point out that the current experimental techniques do not support the simultaneous sampling of the phase-space distribution (but for the possible separate measurements of the density distribution with high-resolution optics~\cite{Bakr2009} or the momentum distribution using the time-of-flight image~\cite{Ketterle1999}).
Our formalism is hopefully verified in experiments once  appropriate techniques become available.
It is emphasized that, although formulated in phase space for the purpose of a straightforward  connection to the Wigner function,  our formalism can be extended to various other spaces, where
the current experimental techniques may be sufficient.

\textit{Nonlinear Fokker-Planck equation incorporating quantum statistics}.---Let us first introduce a %deterministic
nonlinear Fokker-Planck equation~\cite{Kaniadakis2001, Frank2005, Chavanisa2008},
\be
\label{es1}
\frac{\partial \rho_t}{\partial t}=L_\text{st}\rho_t+ \frac{\partial j_t}{\partial p},
\ee
with a flux density in phase space $j_t$ originating from collisions between the system and the reservoir. It reads
\be
\label{ej}
j_t=\gamma  p\rho_t(1+\epsilon\rho_t) +\gamma m k_\text{B} T\frac{\partial \rho_t}{\partial p},
\ee
where $L_\text{st}=-\frac{p}{m}\frac{\partial }{\partial x}+\frac{\partial U}{\partial x}\frac{\partial }{\partial p}$ denotes the streaming operator,
%\be
%L_\text{st}=-\frac{p}{m}\frac{\partial }{\partial x}+\frac{\partial U}{\partial x}\frac{\partial }{\partial p},
%\ee
$m$ the mass of the particle, $\gamma$ the damping coefficient, $U$ the potential energy of the particle, $k_\text{B}$ the Boltzmann constant, and $T$ the temperature of the reservoir. Here, $\epsilon =1, -1, 0$ for  bosons, fermions  and distinguishable  particles, respectively. In the derivation, we have used the Kramers–Moyal expansion of the master equation and truncated the expansion up to the second order~\cite{Kaniadakis2001, Risken1996}.

%For noninteracting spinless particles satisfying quantum statistics, $\epsilon =1, -1, 0$ for  bosons, fermions  and distinguishable  particles respectively.
For bosons, Eq.~(\ref{es1}) describes the evolution of particles above the critical temperature of the Bose-Einstein condensation. %or the noncondensate components below the critical temperature. The condensed components are not considered in this paper.
The Bose-Einstein condensation is not considered in this paper.
For fermions, due to the Pauli exclusion principle, we require $0\leq\rho_t\leq1$~\cite{foot3}.

Equation~(\ref{es1}) is  conservative in particle numbers, $N[\rho_t]=\int \mathrm dz \rho_t(z)$ %, i.e.,$\int \mathrm dz \rho_t(z)=N$, where $\mathrm dz=\mathrm dx\mathrm dp/h$, $h$ denotes the Planck constant.
%\be
%\label{e2}
%N[\phi]=\int \mathrm dz \rho_t(z),
%\ee
where $\mathrm dz=\mathrm dx\mathrm dp/h$, and $h$ denotes the Planck constant.
Also, it determines a steady state (a semiclassical equilibrium state in phase space)
\be
\label{e3}
\rho_{\text{eq}}(z)=\frac{1}{\mathrm e^{\beta\left[p^2/(2m)+U(x)-\mu\right]}-\epsilon},
\ee
where $\beta=1/(k_\text{B} T)$ is the inverse temperature, and $\mu$   the chemical potential. Equation~(\ref{e3}) can be found in Refs.~\cite{Giorgini1999, Zaremba1999, Nikuni1999} for bosons by ignoring the two-body interaction, and  in Ref.~\cite{Nikiforov2005}  for fermions.

To highlight the connection between Eq.~(\ref{es1}) and thermodynamics, we rewrite it in the following form (also see Ref.~\cite{Chavanisa2008}),
\be
\label{e4}
\frac{\partial \rho_t}{\partial t}=L_\text{st}\rho_t+ L_{\text{fp}}(\rho_t)\frac{\delta F[\rho_t]}{\delta \rho_t},
\ee
where the operator %$L_{\text{fp}}$ reads
$L_{\text{fp}}(\rho)=m\gamma\frac{\partial }{\partial p}\left[\rho(1+\epsilon\rho)\frac{\partial}{\partial p}\right]$. %(given $D=m\gamma k_\text{B}T$ as the diffusion constant)
%\be
%\label{efp}
%L_{\text{fp}}(\rho)=m\gamma\frac{\partial }{\partial p}\left[\rho(1+\epsilon\rho)\frac{\partial}{\partial p}\right].
%\ee
Here for the system, %with respect to $\rho_t$,
$F[\rho_t]=E[\rho_t]-TS[\rho_t]$ is the free energy,  where the internal energy $E[\rho_t]$ and the entropy $S[\rho_t]$ are respectively given by
\be
\label{eee}
E[\rho_t]=\int\mathrm dz\rho_t\left(\frac{p^2}{2m}+U\right),
\ee
and
\be
\label{ees}
S[\rho_t]=k_\text{B}\int\mathrm dz[-\rho_t\ln \rho_t+\epsilon^{-1}(1+\epsilon \rho_t)\ln (1+\epsilon \rho_t)].
\ee

\textit{Stochastic  Fokker-Planck equation}.---
It seems straightforward to construct a Langevin equation for a single particle corresponding to the nonlinear Fokker-Planck equation~(\ref{es1}) (see examples in Sec. 6. 5. 4 of Ref.~\cite{Frank2005}), and establish stochastic thermodynamics on that. However, it is inconsistent to incorporate quantum statistics in a Langevin equation for a single particle. This is because that quantum statistics is an effect of the exchange symmetry of a many-body system. In contrast, we  introduce fluctuations of the many-body system in another manner:  adding a noise term in the flux density $j_t$ (Eq.~\ref{ej}).

%Now, let us consider the finite-$N$ effect of the particle-number flux $j_t$ (Eq.~(\ref{ej})).
Due to the discreteness of  particle numbers and the randomness of   collisions between the system and the reservoir, Eq.~(\ref{es1}) only describes the evolution of $\rho_t$ on average for a finite-size system. Hence, we add a noise term $\eta_t$ into $j_t$ to characterize the finite-$N$ effects of the dynamics. %study the evolution of a many-body system with thermal fluctuation by adding a noise term into Eq.~(\ref{es1}).
The introduction of the noise term in the position-space flux density has been used to study many systems, such as  many-body Brownian motion~\cite{Dean1996},  and hydrodynamic fluctuations~\cite{Landau2013}. In particular, it is used to describe the turbulence in randomly stirred fluids~\cite{Forster1976},  the onset of hydrodynamic stabilities~\cite{Hohenberg1992}, and charge transport in semiconducting devices~\cite{Gu2018}. Moreover,  the connection between the stochastic Fokker-Planck equation (Eq.~\ref{e7}) and the Bogoliubov-Born-Green-Kirkwood-Yvon hierarchy is discussed in Ref.~\cite{Chavanis20082}.

According to the above discussions,  we  postulate a stochastic  Fokker-Planck equation (similar results for density in position space are shown in Refs.~\cite{Chavanis2011, Chavanis2019}): %which describes the evolution of the many-body system with finite-$N$ effects
%\begin{widetext}
\be
\label{e7}
\frac{\partial \rho^N_t}{\partial t}=L_{\text{st}}\rho_t^N+L_{\text{fp}}(\rho_t^N)\frac{\delta F[\rho^N_t]}{\delta \rho^N_t}+\frac{\partial \eta_t}{\partial p},
\ee
%\end{widetext}
where $\eta_t$ is a %phase-space-time
Gaussian white noise satisfying $\langle\eta_t(z)\rangle=0$, $\langle\eta_t(z)\eta_{t'}(z')\rangle=2h m\gamma k_\text{B}T\rho^N_t(z)[1+\epsilon \rho^N_t(z)]\delta(z-z')\delta(t-t')$. %, with $D=m\gamma k_\text{B}T$ denotes the diffusion constant.
Such an equation is still conservative in particle number.
%And the derivative ahead of $\eta$ still preserves the conservation of particle number.
%The subscript $N$ emphasizes that $\rho_t^N$ is a random variable due to finite-$N$ effects. The derivative ahead of $\eta$ still preserves the conservation of particle number. Because $\rho^N_t\sim N$, we have $\eta/\rho^N_t\sim1/\sqrt{N}$~\cite{}. Thus,
In the thermodynamic limit, the suppression of the noise $\eta_t$  is shown in Refs.~\cite{Chavanis2011,  Chavanis2019}.

Since the  phase-space distribution $\rho_t^N$ now is  a random variable, we define the probability distribution of observing a fixed distribution $\phi$  at time $t$ as $P_t[\phi]=\langle\prod_{z}\delta[\phi(z)-\rho^N_t(z)]\rangle$. Then, following the similar procedure in Refs.~\cite{Chavanis2011,  Chavanis2019, Frusawa2000,  Goldenfeld1992}, its evolution equation (the functional Fokker-Planck equation) is given by
\begin{widetext}
\be
\label{e8}
\frac{\partial P_t[\phi]}{\partial t}=-\int\mathrm dz \frac{\delta}{\delta \phi(z)}\left\{P_t[\phi]L_\text{st}\phi(z)+L_{\text{fp}}(\phi(z))\left[\frac{\delta F[\phi]}{\delta \phi(z)}P_t[\phi]+k_\text{B} T\frac{\delta P_t[\phi]}{\delta \phi(z)}\right]\right\}.
\ee
\end{widetext}
Its stationary solution is
\be
\label{e9}
P_\text{s}[\phi]=\mathcal Z^{-1}\mathrm e^{-\beta F[\phi]}\delta(N-N[\phi]),
\ee
where $\mathcal Z=\int \mathcal{D}\phi \mathrm e^{-\beta F[\phi]}\delta(N-N[\phi])$ is the generalized partition function. %$N[\phi]$ is given in Eq.~(\ref{e2}), and
%\be
%\mathcal Z=\int \mathcal{D}\phi e^{-\beta F[\phi]}\delta(N-N[\phi])\delta(L_\text{st}\phi).
%\ee
Also, the integral $\int \mathcal{D}\phi$ is constrained by the condition: $\phi\geq 0$ ($0\leq\phi\leq1$ for fermions). The stationary solution is actually a quasi-equilibrium state in a canonical system according to the theory of equilibrium fluctuations~\cite{Mishin2015}, which precisely determines the form of the noise term~\cite{Chavanis2019}. In the thermodynamic limit, $\phi$ converges to $\rho_{\text{eq}}$ in probability by using the method of steepest descent.
%we have  $P_\text{s}[\phi]=\prod_{z}\delta[\phi(z)-\rho_{\text{eq}}(z)]$
In other words, according to the minimum free-energy principle in a canonical ensemble, $P_\text{s}[\phi]$ now is replaced by its  most probable distribution $\rho_{\text{eq}}$. %is the most probable distribution of  $P_\text{s}[\phi]$.

The functional Fokker-Planck equation has an equivalent path-integral form. Let us define a trajectory of the stochastic phase-space distribution as $\phi_{[0,t]}:=\{\phi_s|s\in[0,t]\}$. Then, the   probability distribution of the trajectory conditioned with a fixed distribution $\phi_0$ at initial time $t=0$ reads $P[\phi_{[0,t]}|\phi_0]\propto \mathrm e^{- \beta\mathcal S[\phi_{[0,t]}]}$,
%\be
%P[\phi_{[0,t]}|\phi_0]=\frac{ e^{- \beta\mathcal S[\phi_{[0,t]}]}}{\int \mathcal D\phi_t\int_{\phi_0}^{\phi_t}\mathcal D\phi_{[0,t]} e^{- \beta\mathcal S[\phi_{[0,t]}]}}
%\ee
where the action $\mathcal S$ is a generalized Onsager-Machlup functional,
\begin{widetext}
 \begin{gather}
  \begin{split}
  \label{e18}
\mathcal S[\phi_{[0,t]}]=-\frac{1}{4}\int_{0}^{t}\mathrm ds\int\mathrm dz\left[\frac{\partial \phi_s}{\partial s}-L_{\text{st}}\phi_s-L_{\text{fp}}(\phi_s)\frac{\delta F[\phi_s]}{\delta \phi_s}\right]L_\text{fp}(\phi_s)^{-1}\left[\frac{\partial \phi_s}{\partial s}-L_{\text{st}}\phi_s-L_{\text{fp}}(\phi_s)\frac{\delta F[\phi_s]}{\delta \phi_s}\right].
    \end{split}
 \end{gather}
\end{widetext}
Similar results can be found  in Refs.~\cite{Chavanis2011, Chavanis2019, Tauber1992, Onsager1953}.
Here,  the expression $L_\text{fp}(\phi)^{-1}$ should be
understood as the Green's function of $L_\text{fp}(\phi)$.
According to the principle of least action (the steepest descent approximation of $\mathcal S[\phi_{[0,t]}]$), the dynamics of $\phi_t$ converges to the nonlinear Fokker-Planck equation~(\ref{e4}) in probability in the thermodynamic limit~\cite{foot2}. %i. e.,

\textit{Stochastic thermodynamics}.---When the potential energy is tuned by a time-dependent parameter $\lambda_t$, we can establish the stochastic thermodynamics incorporating quantum statistics along a given trajectory of the phase-space distribution $\phi_{[0, t]} $. %and the corresponding $N$-particle microscopic interpretation $z_i(t)=(x_i(t),p_i(t))$:

%The stochastic internal energy with the explicit $\lambda_t$ dependence is $e[\phi_t; \lambda_t]=E[\phi_t; \lambda_t]$ (Eq.~\ref{eee}).

The stochastic work $w[\phi_{[0, t]}]$, and the stochastic heat $q[\phi_{[0, t]}]$ are defined as
%\be
%e[\phi_t]=E[\phi_t],%=\sum_{i=1}^{N}\left[\frac{p_i(t)^2}{2m}+U(x_i(t),\lambda_t)\right],
%\ee
 \begin{gather}
  \begin{split}
  \label{ew}
w[\phi_{[0, t]}]=&\int_{0}^{t}\mathrm ds\int\mathrm dz\phi_s(z)\frac{\mathrm d\lambda_s}{\mathrm ds}\frac{\partial  U(x,\lambda_s)}{\partial \lambda_s},%\\
%=&\sum_{i=1}^{N}\int_{0}^{t}\mathrm ds \frac{d\lambda_s}{ds} \frac{\partial U(x_i(s),\lambda_s)}{\partial \lambda_s},
    \end{split}
 \end{gather}
and
 \begin{gather}
  \begin{split}
  \label{eq}
q[\phi_{[0, t]}]=&\int_{0}^{t}\mathrm ds\int\mathrm dz\frac{\partial \phi_s(z)}{\partial s}\left[\frac{p^2}{2m}+U(x,\lambda_s)\right].%\\
%=&\sum_{i=1}^{N}\left[ \frac{p_i(t)^2-p_i(0)^2}{2m}+\int_{0}^{t}\mathrm ds \frac{\partial U(x_i(s),\lambda_s)}{\partial x_i(s)}p_i(s)\right].
    \end{split}
 \end{gather}
%When $N=1$, we reproduce the conventional definition of stochastic internal energy, stochastic work, and stochastic heat~\cite{Sekimoto2010,  Seifert2012, Klages2013}.
They satisfy the conservation law of energy
\be
E[\phi_t; \lambda_t]-E[\phi_0; \lambda_0]=w[\phi_{[0, t]}]+q[\phi_{[0, t]}].
\ee
It is worth mentioning that for a deterministic trajectory of $\phi_s$, the definitions of work~(\ref{ew}) and heat~(\ref{eq}) are the semiclassical phase-space counterparts of the quantum work and the quantum heat given in Ref.~\cite{quan2007}.

The stochastic entropy $s[\phi_t]$, stochastic total entropy production $s_{\text{p}}[\phi_{[0, t]}]$, and stochastic free energy, $f[\phi_t; \lambda_t]$ are respectively given by
 \begin{gather}
  \begin{split}
  \label{es}
s[\phi_t]=S[\phi_t]-k_\text{B}\ln P_t[\phi_t],
%=-k_\text{B}\ln P_t[\phi_t]+\sum_{i=1}^{N}\left[- k_\text{B}\ln \phi_t(z_i(t))+k_\text{B}[1+\epsilon^{-1}\phi_t(z_i(t))^{-1}]\ln [1+\epsilon \phi_t(z_i(t))],
    \end{split}
 \end{gather}
\be
\label{e27}
s_{\text{p}}[\phi_{[0, t]}]=s[\phi_t]-s[\phi_0]+s_\text{r}[\phi_{[0, t]}],
\ee
\be
\label{e28}
f[\phi_t; \lambda_t]=E[\phi_t; \lambda_t]-Ts[\phi_t]=F[\phi_t; \lambda_t]+k_\text{B}T\ln P_t[\phi_t],
\ee
where $P_t[\phi]$ is the solution  of the functional Fokker-Planck equation~(\ref{e8}), and
$s_\text{r}[\phi_{[0, t]}]=-q[\phi_{[0, t]}]/T$ denotes the entropy change of the reservoir.

While the stochastic internal energy $E[\phi_t; \lambda_t]$, the stochastic work $w[\phi_{[0, t]}]$, and the stochastic heat $q[\phi_{[0, t]}]$ are solely determined by $\phi_t$, Eq.~(\ref{es}) means that  the stochastic entropy $s[\phi_t]$ also depends on the probability distribution of the stochastic phase-space distribution $P_t[\phi]$. That is to say, the explicit expression of $s[\phi_t]$ cannot be determined until we have obtained the solution of the  functional Fokker-Planck equation. Moreover, if we take an average of $s[\phi_t]$ over $P_t[\phi]$, we have
\be
\langle s[\phi_t]\rangle=\langle S[\phi_t]\rangle-k_\text{B}\int \mathcal D \phi P_t[\phi]\ln P_t[\phi].
\ee
The term $-k_\text{B}\ln P_t[\phi_t]$ actually has a contribution to the average entropy due to the probability distribution $P_t[\phi]$ from the perspective of information theory~\cite{Jaynes1957, Jaynes1983}.
%When $N=1$, $\epsilon=0$,  $\phi_t$ is replaced by $\rho_t$, and $\ln P_t[\rho_t]=0$,  we have $s[\phi_t]=- k_\text{B}\sum_{i=1}^{N} \ln \phi_t(z_i(t))$, which reproduces the conventional definition of stochastic entropy~\cite{Seifert2012, Klages2013}.
Also, this term is essential to prove %For the entropy production $s_{\text{p}}$ (Eq.~(\ref{e27})), using
%integral by parts, we obtain
the $H$ theorem $\frac{\mathrm d}{\mathrm dt}\langle s_{\text{p}}[\phi_{[0, t]}]\rangle\geq0$ (see Appendix A).

We prove the fluctuation theorems according to these stochastic thermodynamic quantities. For this purpose, let $ P[\phi_{[0,t]}|\phi_0] $ ($ P^\dag[\phi^{ \dag}_{[0,t]}|\phi^{ \dag}_0] $) denote the conditional probability distribution of the forward (reverse) trajectory $\phi_{[0, t]} $ ($\phi_{[0, t]}^{ \dag}:=\{\phi^\dag_s|s\in[0,t]\},  \phi^\dag_s(x, p):=\phi_{t-s}(x, -p)$) with a fixed initial distribution $\phi_0$ ($\phi^{ \dag}_0$) and a  forward (reverse) protocol $\lambda_s$ ($\lambda^\dag _s:=\lambda_{t-s}$). Then, using the generalized Onsager-Machlup functional (Eq.~\ref{e18}), we obtain the detailed fluctuation theorems  (see Appendix C):
 \begin{gather}
  \begin{split}
\ln\frac{P[\phi_{[0,t]}|\phi_0]}{P^\dag[\phi^{ \dag}_{[0,t]}|\phi^{ \dag}_0]}=\ln\frac{P_t[\phi_t]}{P_0[\phi_0]}+\frac{s_\text{p}[\phi_{[0, t]}]}{k_\text{B}}.
    \end{split}
 \end{gather}
% \begin{gather}
%  \begin{split}
%\ln\frac{P[\phi_{[0,t]}|\phi_0]}{P^\dag[\phi^{ \dag}_{[0,t]}|\phi^{ \dag}_0]}=& -\beta\int_{0}^{t}\mathrm ds\int\mathrm dz \frac{\partial \phi_{s}}{\partial t} \frac{\delta F[\phi_s]}{\delta \phi_s} \\
%=& -\beta(F[\phi_t]-F[\phi_0]-w[\phi_{[0,t]}])\\
%=& (S[\phi_t]-S[\phi_0]+s_\text{r}[\phi_{[0, t]}])/k_\text{B}\\
%=&\ln\frac{P_t[\phi_t]}{P_0[\phi_0]}+s_\text{p}[\phi_{[0, t]}]/k_\text{B}.
%    \end{split}
% \end{gather}
By adding an arbitrary normalized distribution of $\phi$ at the  initial time of the forward (reverse) process $P_0[\phi_0]$ ($P'_0[\phi_0^\dag]$), we obtain the integral fluctuation theorems:
\be
\left\langle  \frac{P'_0[\phi_0^\dag]}{P_t[\phi_t]} \mathrm e^{-s_\text{p} /k_\text{B}}\right\rangle=1.
\ee
Such an equality is formally consistent with the integral fluctuation theorems in previous studies~\cite{Seifert2012, Klages2013}.

For a choice of  $P'_0[\phi^\dag_0]=P_t[\phi_t]$, we obtain the integral fluctuation theorem for total entropy production~\cite{Seifert2012, Klages2013},
\be
\left\langle    \mathrm e^{-s_\text{p} /k_\text{B}}\right\rangle=1.
\ee
%\be
%\label{e31}
%\left\langle    e^{-s_\text{p} /k_\text{B}}\right\rangle=1.
%\ee
Then as a corollary, the second law $\left\langle s_\text{p} \right\rangle\geq 0$ follows from the fluctuation theorem   by using Jensen's inequality.

Moreover, when both $P_0[\phi_0], P'_0[\phi_0^\dag]$  are stationary solutions of the functional Fokker-Planck equation (Eq.~\ref{e9}), i. e., $P_0[\phi_0]=P_\text{s}[\phi_0; \lambda_0], P'_0[\phi_0^\dag]=P_\text{s}[\phi_t; \lambda_t]$, we obtain the generalized Jarzynski equality
\be
\left\langle   \mathrm  e^{-\beta w }\right\rangle= \frac{\mathcal Z(\lambda_t)}{\mathcal Z(\lambda_0)},
\ee
%\be
%\label{e32}
%\left\langle    e^{-\beta w }\right\rangle= \frac{\mathcal Z(\lambda_t)}{\mathcal Z(\lambda_0)},
%\ee
and the generalized principle of maximum work
\be
\langle w \rangle \geq -k_\text{B}T\ln\left[ \frac{\mathcal Z(\lambda_t)}{\mathcal Z(\lambda_0)}\right]
\ee
by using Jensen's inequality.
In the limit $N\to\infty, \mathcal Z(\lambda_{t})/\mathcal Z(\lambda_{0})=\mathrm e^{-\beta(F[\rho_\text{eq}(\lambda_t)]-F[\rho_\text{eq}(\lambda_0)])}$ by using the steepest descent approximation of $\mathcal Z$, %as $P_\text{s}[\phi]$ is replaced by its  most probable distribution $\rho_{\text{eq}}$ here,
which further results in the Jarzynski equality $\langle \mathrm e^{-\beta w} \rangle = \mathrm e^{-\beta(F[\rho_\text{eq}(\lambda_t)]-F[\rho_\text{eq}(\lambda_0)])}$,  and the  principle of maximum work $\langle w \rangle \geq F[\rho_\text{eq}(\lambda_t)]-F[\rho_\text{eq}(\lambda_0)]$~\cite{Sekimoto2010,  Seifert2012, Klages2013}.

We would like to emphasize that for the stationary solution $P_\text{s}[\phi;\lambda]$~(\ref{e9}), we have
\be
\langle f[\phi; \lambda]\rangle=-k_{\text{B}}T\ln \mathcal Z(\lambda),
\ee
\be
\langle E[\phi; \lambda]\rangle=-\frac{\partial\ln \mathcal Z(\lambda)}{\partial \beta},
\ee
\be
 \frac{\langle s[\phi; \lambda]\rangle}{k_{\text{B}}}= \ln \mathcal Z(\lambda) -\beta\frac{\partial\ln \mathcal Z(\lambda)}{\partial \beta} .
\ee
Hence, the generalized partition function $\mathcal Z(\lambda)$ as the characteristic state function of a canonical quasi-equilibrium state plays the same role as the partition function of a canonical equilibrium state.

For distinguishable particles ($\epsilon=0$), our formalism coincides with previous stochastic thermodynamics. It is demonstrated by following Dean's study of a system of $N$-particle Langevin equations~\cite{Dean1996},
 \begin{gather}
  \begin{split}
\frac{\mathrm d x_i(t)}{\mathrm dt}=&\frac{p_i(t)}{m},\\
\frac{\mathrm d p_i(t)}{\mathrm dt}=&-\frac{\partial U(x_i(t))}{\partial x_i(t)}-\gamma p_i(t)+\xi_i(t),
    \end{split}
 \end{gather}
 %where the phase-space density function $\tilde{\rho}^N_t(z)=\sum_{i=1}^{N}\delta(z-z_i(t))$ in the self-consistent damping %$\gamma(\tilde{\rho}_t^N)=\gamma(1+\epsilon \tilde{\rho}_t^N)$
%originates from the surrounding particles due to quantum statistics.
where  the white noises are uncorrelated, $\langle\xi_i(t)\rangle=0$, $\langle\xi_i(t)\xi_j(t')\rangle=2m\gamma k_\text{B}T\delta_{ij}\delta(t-t')$. Then, the empirical density $\tilde{\rho}^N_t(z)=\sum_{i=1}^{N}\delta(z-z_i(t))$ of particles satisfies the stochastic Fokker-Planck equation~(\ref{e7}), where $\eta_t=\sum_{i=1}^{N}\xi_i(t)$ correspondingly. Substituting $\tilde{\rho}^N_t(z)$ into the definition of stochastic thermodynamic quantities $E[\phi_t; \lambda_t]$, $w[\phi_{[0, t]}]$, $q[\phi_{[0, t]}]$, and $s[\phi_t]$ (the second term in Eq.~(\ref{es})),  we reproduce their counterparts in previous studies of  stochastic thermodynamics~\cite{Sekimoto2010,  Seifert2012, Klages2013}. % for noninteracting-particle systems.
%However, the term $-k_\text{B}\ln P_t[\phi_t]$ in Eq.~(\ref{es})  and  the stochastic free energy $f[\phi_t; \lambda_t]$ have not been reported in previous papers, which is essential when we consider the H-theorem and fluctuation theorems for a finite-$N$ system.

\textit{Conclusion}.---For a quantum many-body system semiclassically treated by a phase-space distribution at the mesoscopic level, we consider a stochastic Fokker-Planck equation as the dynamics of the system at the mesoscopic level. Here, the noise term finds its origin in the discreteness of the particle number and the randomness of  collisions between the system and the reservoir, embodying Einstein's interpretation of the reverse form   of the Boltzmann entropy.
%We interpret the stochastic phase-space distribution as the density function of a $N$-particle system in microscopic level satisfying Langevin equations by assuming a specific  correlation between  the thermal noises of different particles.

Based on trajectories of the stochastic phase-space distribution, %and the corresponding $N$-particle microscopic interpretation,
we propose a formalism of stochastic thermodynamics that accounts for quantum statistics. Consequently, a connection between stochastic thermodynamics and the quasi-equilibrium state is established.%, which is different from the connection made in previous studies between stochastic thermodynamics and the equilibrium state.

Independent of the previous formalism that relies on the two-point measurement scheme, our formalism is based on the sampling of the stochastic phase-space distribution, offering hope for experimental verifications. Moreover, by incorporating  mean-field interactions   and considering other types of non-Boltzmann entropy~\cite{Kaniadakis2001, Frank2005, Chavanisa2008}, our formalism can be readily extended to interacting systems and other non-Boltzmann  systems. It is worth mentioning that an efficient treatment of fractional exclusion statistics is given by using the nonlinear Fokker-Planck equation in Ref.~\cite{Kaniadakis1996}.

The author thanks Y. Chen, C. P. Sun,  J. F. Chen, and J. Gu for helpful suggestions.
This work is supported by the NSFC (Grant No. 12088101), NSAF (Grants No. U1930403, and No. U1930402), and  the China Postdoctoral Science
Foundation (No. 2021M700359).

 \renewcommand{\theequation}{A.\arabic{equation}}

 \setcounter{equation}{0}

\begin{widetext}

\section*{Appendix A:  proof of the $H$ theorem}

It follows from Eqs.~(\ref{e27}, \ref{e28}) and the conservation of energy that
 \begin{gather}
  \begin{split}
  \label{e26}
&\frac{\mathrm d}{\mathrm dt}\langle s_{\text{p}}[\phi_{[0, t]}]\rangle\\
%=&\frac{d }{dt}\left(\langle s[\phi_t]\rangle+\langle s_\text{r}[\phi_{[0, t]}]\rangle\right)\\
%=&\frac{d }{dt}\left(\langle s[\phi_t]\rangle-\frac{1}{T}\langle q[\phi_{[0, t]}]\rangle\right)\\
%=&\frac{d }{dt}\left(\langle s[\phi_t]\rangle-\frac{1}{T}\langle e[\phi_t; \lambda_t]\rangle+ \frac{1}{T}\langle w [\phi_{[0, t]}]\rangle\right)\\
=&\frac{1}{T}\frac{\mathrm d }{\mathrm dt}\left( \langle w [\phi_{[0, t]}]\rangle- \langle f[\phi_t; \lambda_t]\rangle \right)\\
=&\frac{1}{T}\frac{\mathrm d }{\mathrm dt}\left( \langle w [\phi_{[0, t]}]\rangle- \langle F[\phi_t; \lambda_t]+k_\text{B}T  \ln P_t[\phi_t]\rangle\right).%\\
%=&-\int\mathcal D\phi \frac{\partial P_t[\phi]}{\partial t}(T^{-1} F[\phi]+k_\text{B}\ln P_t[\phi]+1)\\
%=&\int\mathcal D\phi \int\mathrm dz \frac{\gamma\phi(z)[1+\epsilon \phi(z)]}{TP_t[\phi]}\times\\
% &\left\{\frac{\partial }{\partial p}\left[\frac{\delta F[\phi]}{\delta \phi(z)}P_t[\phi]+k_\text{B} T\frac{\delta P_t[\phi]}{\delta \phi(z)}\right]\right\}^2\\
%\geq &0,
    \end{split}
 \end{gather}
Substituting Eq.~(\ref{ew}) into Eq.~(\ref{e26}), and using Eq.~(\ref{e8}) %, $\langle A[\phi_t(z)]\rangle=\int\mathcal D\phi P_t[\phi]A[\phi(z)]$ for some functional $A$, and $\frac{d\lambda_t}{d t}\frac{\partial F[\phi_t; \lambda_t]}{\partial \lambda_t}=\frac{d\lambda_t}{d t}\frac{\partial E[\phi_t; \lambda_t]}{\partial \lambda_t}=\frac{d }{dt} w [\phi_{[0, t]}]$,
we have (noticing $\frac{\partial  F[\phi; \lambda_t]}{\partial \lambda_t}=\int\mathrm dz \phi(z) \frac{\mathrm d\lambda_t}{\mathrm dt}\frac{\partial  U(x,\lambda_t)}{\partial \lambda_t}$)
 \begin{gather}
  \begin{split}
&\frac{1}{T}\frac{\mathrm d }{\mathrm dt}\left( \langle w [\phi_{[0, t]}]\rangle- \langle F[\phi_t; \lambda_t]+k_\text{B}T  \ln P_t[\phi_t]\rangle\right)\\
=&\frac{1}{T}\left( \int\mathrm dz\langle\phi_t(z)\rangle\frac{\mathrm d\lambda_t}{\mathrm dt}\frac{\partial  U(x,\lambda_t)}{\partial \lambda_t}- \frac{\mathrm d }{\mathrm dt}\langle F[\phi_t; \lambda_t]+k_\text{B}T  \ln P_t[\phi_t]\rangle\right)\\
=&\frac{1}{T}\left( \int\mathcal D\phi P_t[\phi]\int\mathrm dz \phi(z) \frac{\mathrm d\lambda_t}{\mathrm dt}\frac{\partial  U(x,\lambda_t)}{\partial \lambda_t}- \frac{\mathrm d }{\mathrm dt}\langle F[\phi_t; \lambda_t]+k_\text{B}T  \ln P_t[\phi_t]\rangle\right)\\
=&\frac{1}{T}\left(\int\mathcal D\phi P_t[\phi]\frac{\mathrm d\lambda_t}{\mathrm d t}\frac{\partial  F[\phi; \lambda_t]}{\partial \lambda_t}-\frac{\mathrm d }{\mathrm dt}\int\mathcal D\phi P_t[\phi](F[\phi; \lambda_t]+k_\text{B}T  \ln P_t[\phi])\right)\\
=&-\int\mathcal D\phi \frac{\partial P_t[\phi]}{\partial t}(T^{-1} F[\phi; \lambda_t]+k_\text{B}\ln P_t[\phi]+k_\text{B})\\
=&\int\mathcal D\phi \int\mathrm dz \frac{\delta}{\delta \phi(z)}\left\{P_t[\phi]L_\text{st}\phi(z)+L_{\text{fp}}(\phi(z))\left[\frac{\delta F[\phi; \lambda_t]}{\delta \phi(z)}P_t[\phi]+k_\text{B} T\frac{\delta P_t[\phi]}{\delta \phi(z)}\right]\right\}(T^{-1} F[\phi; \lambda_t]+k_\text{B}\ln P_t[\phi]+k_\text{B}).
%=&\int\mathcal D\phi \int\mathrm dz \frac{\gamma\phi(z)[1+\epsilon \phi(z)]}{TP_t[\phi]}\times\\
% &\left\{\frac{\partial }{\partial p}\left[\frac{\delta F[\phi]}{\delta \phi(z)}P_t[\phi]+k_\text{B} T\frac{\delta P_t[\phi]}{\delta \phi(z)}\right]\right\}^2\\
%\geq &0,
    \end{split}
 \end{gather}
%where we have used the relation $\frac{d\lambda_t}{d t}\frac{\partial F[\phi_t; \lambda_t]}{\partial \lambda_t}=\frac{d\lambda_t}{d t}\frac{\partial E[\phi_t; \lambda_t]}{\partial \lambda_t}=\frac{d }{dt} w [\phi_{[0, t]}]$.
Finally, using integral by parts, we obtain
 \begin{gather}
  \begin{split}
  \label{eh}
&\int\mathcal D\phi \int\mathrm dz \frac{\delta}{\delta \phi(z)}\left\{P_t[\phi]L_\text{st}\phi(z)+L_{\text{fp}}(\phi(z))\left[\frac{\delta F[\phi; \lambda_t]}{\delta \phi(z)}P_t[\phi]+k_\text{B} T\frac{\delta P_t[\phi]}{\delta \phi(z)}\right]\right\}(T^{-1} F[\phi; \lambda_t]+k_\text{B}\ln P_t[\phi]+k_\text{B})\\
=&\int\mathcal D\phi P_t[\phi] \int\mathrm dz \left\{-\frac{1}{T}\frac{\delta F[\phi; \lambda_t]}{\delta \phi(z)}+k_\text{B}\frac{\delta }{\delta \phi(z)}\right\}L_\text{st}\phi(z)\\
&+\int\mathcal D\phi \int\mathrm dz \frac{m\gamma\phi(z)[1+\epsilon \phi(z)]}{TP_t[\phi]}\left\{\frac{\partial }{\partial p}\left[\frac{\delta F[\phi; \lambda_t]}{\delta \phi(z)}P_t[\phi]+k_\text{B} T\frac{\delta P_t[\phi]}{\delta \phi(z)}\right]\right\}^2.
% &\left\{\frac{\partial }{\partial p}\left[\frac{\delta F[\phi]}{\delta \phi(z)}P_t[\phi]+k_\text{B} T\frac{\delta P_t[\phi]}{\delta \phi(z)}\right]\right\}^2
%=&\int\mathcal D\phi \int\mathrm dz \frac{\gamma\phi(z)[1+\epsilon \phi(z)]}{TP_t[\phi]}\times\\
% &\left\{\frac{\partial }{\partial p}\left[\frac{\delta F[\phi]}{\delta \phi(z)}P_t[\phi]+k_\text{B} T\frac{\delta P_t[\phi]}{\delta \phi(z)}\right]\right\}^2\\
%\geq &0,
    \end{split}
 \end{gather}
Because the first term  actually equals 0 by using the two identities in Appendix B and the second term is not less than 0, we obtain the $H$ theorem $\frac{\mathrm d}{\mathrm dt}\langle s_{\text{p}}[\phi_{[0, t]}]\rangle\geq0$.

 \renewcommand{\theequation}{B.\arabic{equation}}

 \setcounter{equation}{0}

\section*{Appendix B: two identities in Eq.~(\ref{eh})}

For the first identity, using $F[\phi]=E[\phi]-TS[\phi]$ and integral by parts, we have
 \begin{gather}
  \begin{split}
 &\int\mathrm dz \frac{\delta F[\phi]}{\delta \phi(z)}L_{\text{st}}\phi(z)\\
 =&\int\mathrm dz \left[\frac{p^2}{2m}+U(x,\lambda) \right]L_{\text{st}}\phi(z)-T\int\mathrm dz \frac{\delta S[\phi]}{\delta \phi(z)}L_{\text{st}}\phi(z)\\
 =&-\int\mathrm dz \phi L_{\text{st}}\left[\frac{p^2}{2m}+U(x,\lambda) \right]-k_\text{B}T\int\mathrm dz L_{\text{st}}[-\phi\ln \phi+\epsilon^{-1}(1+\epsilon \phi)\ln (1+\epsilon \phi)]\\
 =&0.
    \end{split}
 \end{gather}

For the second identity, let us first define a set of real orthonormal complete square-integrable functions $\{u_i(z)\}$ in the phase space. Then, we decompose $\phi(z)$ as
\be
\phi(z)=\sum_i \phi_i u_i(z),
\ee
where
\be
\label{a3}
\phi_i=\int dz\phi(z)  u_i(z).
\ee
Hence, it follows from Eq.~(\ref{a3}) that
 \begin{gather}
  \begin{split}
 \int\mathrm dz \frac{\delta}{\delta \phi(z)} L_{\text{st}}\phi(z)
 =&\sum_i\int\mathrm dz \frac{\delta  \phi_i}{\delta \phi(z)} L_{\text{st}} u_i(z)\\
 =&\sum_i\int\mathrm dz  u_i(z) L_{\text{st}} u_i(z)\\
 =&-\sum_i\int\mathrm dz  u_i(z) L_{\text{st}} u_i(z)\\
 =&0,
    \end{split}
 \end{gather}
The minus sign in the third equality  results from integral by parts.

  \renewcommand{\theequation}{C.\arabic{equation}}

 \setcounter{equation}{0}

\section*{Appendix C:  proof of the detailed fluctuation theorem}

We first prove that the normalization factor of the forward propagator $P[\phi_{[0,t]}|\phi_0]$ equals that of the reverse propagator $ P^\dag[\phi^{ \dag}_{[0,t]}|\phi^{ \dag}_0] $.
Let us define another white noise $\zeta_t=\frac{\partial }{\partial p}\eta_t$, where $\langle \zeta_t(z)\rangle=0$ and
 \begin{gather}
  \begin{split}
\langle\zeta_t(z)\zeta_{t'}(z')\rangle=&\frac{\partial^2 }{\partial p \partial p'}\langle\eta_t(z)\eta_{t'}(z')\rangle\\
=&2h k_\text{B}T\frac{\partial }{\partial p }\left\{\rho^N_t(z)[1+\epsilon \rho^N_t(z)]\frac{\partial }{\partial p' }\delta(z-z')\right\}\delta(t-t')\\
=&-2h k_\text{B}T\frac{\partial }{\partial p }\left\{\rho^N_t(z)[1+\epsilon \rho^N_t(z)]\frac{\partial }{\partial p }\delta(z-z')\right\}\delta(t-t')\\
=&-2h k_\text{B}TL_{\text{fp}}(\phi_t(z))\delta(z-z')\delta(t-t').
    \end{split}
 \end{gather}
Thus, %by using the functional Gaussian integral,
the probability distribution of a trajectory $\zeta_{[0,t]}$ reads%~\cite{Tauber1992},
 \begin{gather}
  \begin{split}
P[\zeta_{[0,t]}]=\mathcal N[\phi_{[0,t]}] \mathrm e^{\frac{\beta}{4}\int_0^t\mathrm ds\int \mathrm d z \zeta_s(z)L_{\text{fp}}(\phi_s)^{-1}\zeta_s(z)},
    \end{split}
 \end{gather}
where
\be
\label{en}
\mathcal N[\phi_{[0,t]}]=\int\mathcal D\zeta_{[0,t]}\mathrm e^{\frac{\beta}{4}\int_0^t\mathrm ds\int \mathrm d z \zeta_s(z)L_{\text{fp}}(\phi_s)^{-1}\zeta_s(z)}.
\ee
%$\det[-4\pi k_\text{B}T L_{\text{fp}}(\phi_s)]=\exp[\sum_i\int_0^t\mathrm ds\mathrm dx\ln y_i(\phi_s, x)]$ and $\{y_i[\phi_s, x]\}$ are the eigenvalues of the operator $-4\pi k_\text{B}TL_{\text{fp}}(\phi_s(z))$.

We obtain the propagator $P[\phi_{[0,t]}|\phi_0]$ through
 \begin{gather}
  \begin{split}
&P[\phi_{[0,t]}|\phi_0]\\
=&\int\mathcal D\zeta_{[0,t]} P[\zeta_{[0,t]}]\prod_{(s, z)}\delta\left(\frac{\partial \phi_s(z)}{\partial s}-L_{\text{st}}\phi_s(z)-L_{\text{fp}}(\phi_s(z))\frac{\delta F[\phi_s]}{\delta \phi_s(z)}-\zeta_s(z)\right)\\
=&\mathcal J[\phi_{[0,t]}]\mathcal N[\phi_{[0,t]}]\mathrm e^{- \beta\mathcal S[\phi_{[0,t]}]},
    \end{split}
 \end{gather}
where, the  Jacobian $\mathcal J[\phi_{[0,t]}]$ stemming from the  variable transformation from $\zeta$ to $\phi$ is just a constant  when the standard Ito forward
discretization is applied~\cite{Tauber1992}.

According to Eq.~(\ref{en}) and $\phi^\dag_s(x, p)=\phi_{t-s}(x, -p)$, we have the following equality:
 \begin{gather}
  \begin{split}
  \label{enn}
\mathcal N[\phi^\dag_{[0,t]}]=&\int\mathcal D\zeta_{[0,t]}\mathrm e^{\frac{\beta}{4}\int_0^t\mathrm ds\int \mathrm d z \zeta_s(z)L_{\text{fp}}(\phi^\dag_s)^{-1}\zeta_s(z)}\\
%=&\int\mathcal D\zeta_{[0,t]}e^{\frac{\beta}{4}\int_0^t\mathrm ds\int \mathrm d z \zeta_{t-s}(z)L_{\text{fp}}(\phi^\dag_{t-s})^{-1}\zeta_{t-s}(z)}\\
%=&\int\mathcal D\zeta_{[0,t]}e^{\frac{\beta}{4}\int_0^t\mathrm ds\int \mathrm d z \zeta_{t-s}(x, -p)L_{\text{fp}}(\phi_{s})^{-1}\zeta_{t-s}(x, -p)}\\
=&\int\mathcal D\zeta_{[0,t]}\mathrm e^{\frac{\beta}{4}\int_0^t\mathrm ds\int \mathrm d z \zeta_{s}(z)L_{\text{fp}}(\phi_{s})^{-1}\zeta_{s}(z)}\\
=&\mathcal N[\phi_{[0,t]}],
    \end{split}
 \end{gather}
where we have used the transformation $s\to t-s, p\to -p, \zeta_{t-s}(x, -p)\to \zeta_{s}(x, p)$.

Next, using the transformation $s\to t-s, p\to -p$, it is straightforward to verify that the action $\mathcal S[\phi^\dag_{[0,t]}]$ (Eq.~(\ref{e18})) can be rewritten in terms of $\phi_{[0,t]}$:
\be
\label{c5}
\mathcal S[\phi^\dag_{[0,t]}]=-\frac{1}{4}\int_{0}^{t}\mathrm ds\int\mathrm dz\left[\frac{\partial \phi_s}{\partial s}-L_{\text{st}}\phi_s+L_{\text{fp}}(\phi_s)\frac{\delta F[\phi_s]}{\delta \phi_s}\right]L_\text{fp}(\phi_s)^{-1}\left[\frac{\partial \phi_s}{\partial s}-L_{\text{st}}\phi_s+L_{\text{fp}}(\phi_s)\frac{\delta F[\phi_s]}{\delta \phi_s}\right].
\ee
%Because $\det[-4\pi k_\text{B}T L_{\text{fp}}(\phi_{[0,t]})]=\det[-4\pi k_\text{B}T L_{\text{fp}}(\phi^\dag_{[0,t]})]$ and

Finally, we derive the detailed fluctuation theorem: %using $\frac{d\lambda_t}{d t}\frac{\partial F[\phi_t; \lambda_t]}{\partial \lambda_t}=\frac{d\lambda_t}{d t}\frac{\partial E[\phi_t; \lambda_t]}{\partial \lambda_t}=\frac{d }{dt} w [\phi_{[0, t]}]$, it follows from  Eqs.~(\ref{e18}, \ref{ew}, \ref{e27}, \ref{e28}, \ref{enn}, \ref{c5}) that %(noticing the time-reversal symmetry of each term in Eq.~(\ref{e18}))
 \begin{gather}
  \begin{split}
\ln\frac{P[\phi_{[0,t]}|\phi_0]}{P^\dag[\phi^{ \dag}_{[0,t]}|\phi^{ \dag}_0]}=&-\beta (\mathcal S[\phi_{[0,t]}]-\mathcal S[\phi^\dag_{[0,t]}])\\
=& -\beta\int_{0}^{t}\mathrm ds\int\mathrm dz \frac{\partial \phi_{s}}{\partial s} \frac{\delta F[\phi_s; \lambda_s]}{\delta \phi_s} \\
=& -\beta(F[\phi_t; \lambda_t]-F[\phi_0; \lambda_0]-w[\phi_{[0,t]}])\\
%=& (S[\phi_t]-S[\phi_0]+s_\text{r}[\phi_{[0, t]}])/k_\text{B}\\
=&\ln\frac{P_t[\phi_t]}{P_0[\phi_0]}+\frac{s_\text{p}[\phi_{[0, t]}]}{k_\text{B}},
    \end{split}
 \end{gather}
\end{widetext}
where in the first line we have used Eq.~(\ref{enn}), in the second line we have used Eqs.~(\ref{e18},  \ref{c5}), in the third line we have used $\frac{\mathrm d\lambda_t}{\mathrm d t}\frac{\partial F[\phi_t; \lambda_t]}{\partial \lambda_t}=\frac{\mathrm d\lambda_t}{\mathrm d t}\frac{\partial E[\phi_t; \lambda_t]}{\partial \lambda_t}=\frac{\mathrm d }{\mathrm dt} w [\phi_{[0, t]}]$, and in the fourth line we have used the definitions of the stochastic thermodynamic quantities (Eqs.~(\ref{ew}-\ref{e28})).


\begin{thebibliography}{99}

\bibitem{Sekimoto2010} K. Sekimoto, \emph{Stochastic energetics} (Springer, Berlin, 2010).
%\bibitem{Jarzynski2011} C. Jarzynski, Annu. Rev. Condens. Matter Phys. \textbf{2}(1), 329-351 (2011).
\bibitem{Seifert2012} U. Seifert, Reports on progress in physics \textbf{75}(12), 126001 (2012).
\bibitem{Klages2013}  \emph{Nonequilibrium statistical physics of small systems}, edited by R. Klages, W. Just, and C. Jarzynski (Wiley-VCH, Weinheim, Ger., 2013).
\bibitem{Crooks1999} G. E. Crooks,  Phys. Rev. E 60(3), 2721 (1999); G. Hummer, and A. Szabo, Proc. Natl. Acad. Sci. U. S. A. \textbf{98}(7), 3658-3661 (2001).



\bibitem{aq2000} J. Kurchan, arXiv: cond-mat/0007360; H. Tasaki, arXiv: cond-mat/0009244.
\bibitem{Talkner1999} P. Talkner, E. Lutz and P. H\"{a}nggi, Phys. Rev. E \textbf{75}, 050102(R) (2007).

\bibitem{Hekking2013} F. W. J. Hekking and J. P. Pekola, Phys. Rev. Lett. \textbf{111}, 093602 (2013); M. Esposito, U. Harbola, and S. Mukamel, Rev. Mod. Phys. \textbf{81}, 1665 (2009); J. M. Horowitz and J. M. Parrondo, New J. Phys. \textbf{15}, 085028 (2013); F. Liu, Prog. Phys. \textbf{38}, 1 (2018).
\bibitem{Funo2018} K. Funo, H. T. Quan, Phys. Rev. Lett. \textbf{121}, 040602 (2018); Phys. Rev. E \textbf{98}, 012113 (2018).
\bibitem{Strasberg2021} P. Strasberg, \emph{Quantum Stochastic Thermodynamics: Foundations and Selected Applications} (Oxford University Press, Oxford, U. K., 2022).

\bibitem{foot1} The phase-space distribution is defined by using the Wigner function of the one-particle density matrix [see Eqs.~(17, 18) in Ref.~\cite{Zaremba1999}].


\bibitem{Pathria2011} R. K. Pathria, and P. D. Beale, \emph{Statistical Mechanics}, 3rd ed. (Academic Press, Cambridge, MA, 2011).
\bibitem{Kaniadakis2001} G. Kaniadakis, Physica A \text{296},  405–425 (2001); Physics Letters A \text{288}, 283–291 (2001).
\bibitem{Frank2005}  T. D. Frank, \emph{Nonlinear Fokker–Planck Equations: Fundamentals and Applications} ( Springer, Berlin, 2005).
\bibitem{Chavanisa2008} P. H. Chavanisa, Eur. Phys. J. B \textbf{62}, 179–208 (2008).

\bibitem{Kadanoff2000} L. P. Kadanoff, \emph{Statistical physics: statics, dynamics and renormalization} (World Scientific, Singapore, 2000).
\bibitem{Wolschin1982} G. Wolschin,  Phys. Rev. Lett. \textbf{48}(15), 1004 (1982).
\bibitem{Wolschin2018} G. Wolschin,  Physica A \textbf{499}, 1-10 (2018).
\bibitem{Daligault2016} P. Danielewicz,  Physica A  \textbf{100}(1), 167-182 (1980); J. Daligault, Physics of Plasmas, \textbf{23}(3), 032706 (2016).

\bibitem{Mishin2015} Y. Mishin, Annals of Physics \textbf{363}, 48–97 (2015).
\bibitem{Einstein1910} A. Einstein, Ann. Phys., Lpz. \textbf{33}  1275–1298 (1910).

\bibitem{Bakr2009} W. S. Bakr, J. I. Gillen, A. Peng, S. Föling, and M. Greiner, Nature \textbf{462}, 74-77 (2009); L. W. Cheuk, M. A. Nichols, M. Okan, T. Gersdorf, V. V. Ramasesh, W. S. Bakr, T. Lompe, and M. W. Zwierlein, Phys. Rev. Lett. \textbf{114}, 193001 (2015).
\bibitem{Ketterle1999} W. Ketterle, D. S. Durfee, and D. M. Stamper-Kurn, in \emph{Bose-Einstein condensation in atomic gases}, Proceedings of the International School of Physics ``Enrico Fermi'', Course CXL, Varenna, 1998, edited by M. Inguscio, S. Stringari, and C. E. Wieman (IOS Press, Amsterdam, 1999), p. 67.
\bibitem{Risken1996} H. Risken,  \emph{The Fokker-Planck Equation}  (Springer, Berlin, 1996).
\bibitem{foot3} With a proper initial distribution $0\leq\rho_0\leq1$, the phase-space density $\rho_t$ remains bounded by $0\leq\rho_t\leq1$ during the whole evolution. It is proved by following the similar procedure in  Ref.~\cite{Risken1996} for $\rho_t\geq 0$ and in Ref.~\cite{Carrillo2008} for $\rho_t\leq 1$.
\bibitem{Carrillo2008} J. A. Carrillo, J. Rosado, F. Salvarani, Appl. Math. Lett. \textbf{21},  148–154 (2008).

\bibitem{Zaremba1999} E. Zaremba, T. Nikuni, and A. Griffin, J. Low Temp. Phys. \textbf{116}, 277 (1999).
\bibitem{Giorgini1999} S. Giorgini,  L. P. Pitaevskii,  and S. Stringari, J. Low Temp. Phys. \textbf{109}, 309 (1997).

\bibitem{Nikuni1999} T. Nikuni, E. Zaremba, and A. Griffin, Phys. Rev. Lett. \textbf{83}, 10 (1999).
\bibitem{Nikiforov2005} A. F. Nikiforov,V. G.  Novikov, V. B. Uvarov,  \emph{Quantum-Statistical Models of Hot Dense Matter: Methods for Computation Opacity and Equation of State} (Springer, Berlin, 2005).







\bibitem{Dean1996} D. S. Dean,   J. Phys. A Math. Gen.  \textbf{29}, L613–L617, (1996).
\bibitem{Landau2013} L. D. Landau,   E. M. Lifshitz,  \emph{Fluid Mechanics}, Course of Theoretical Physics, Vol. 6 (Elsevier, Amsterdam, 2013).
\bibitem{Forster1976} D. Forster, D. R. Nelson and M. J. Stephen, Phys. Rev. Lett. \textbf{36}, 867 (1976); Phys. Rev. A \textbf{16}, 732 (1977).
\bibitem{Hohenberg1992} P. C. Hohenberg and J. B. Swift,  Phys. Rev. A \textbf{46}, 4773 (1992); J. B. Swift, K. L. Babcock and P. C. Hohenberg, Physica A \textbf{204}, 625 (1994).
\bibitem{Gu2018} J. Gu, and P. Gaspard,  Phys. Rev. E \textbf{97}, 052138 (2018);   \textbf{99}, 012137 (2019).
\bibitem{Chavanis20082} P. H. Chavanis, Physica A \textbf{387},  5716–5740 (2008).


\bibitem{Chavanis2011} P. H. Chavanis,   Physica A  \textbf{390}, 1546–1574 (2011);  Entropy \textbf{17}, 3205 (2015); P. H. Chavanis,  and L. Delfini,  Phys. Rev. E \textbf{89}, 032139 (2014).
\bibitem{Chavanis2019} P. H. Chavanis, Entropy  \textbf{21}, 1006  (2019).
\bibitem{Frusawa2000} H. Frusawa, and R. Hayakawa, J. Phys. A: Math. Gen. \textbf{33},  L155–L160 (2000).
%\bibitem{Justin2002} J. Zinn-Justin,  \emph{Quantum Field Theory and Critical Phenomena} (Oxford: Oxford University Press, 2002) ch 4.
\bibitem{Goldenfeld1992} N. Goldenfeld,  \emph{Lectures on Phase Transitions and the Renormalization Group} ( Addison-Wesley, Reading, MA, 1992).
\bibitem{Tauber1992} U. C. T\"{a}uber , \emph{Critical dynamics: a field theory approach to equilibrium and nonequilibrium scaling behavior} (Cambridge University Press, Cambridge, U. K., 2014).
\bibitem{Onsager1953} L. Onsager, S. Machlup,    Phys. Rev.  \textbf{91}, 1505–1512 (1953).
\bibitem{foot2} This conclusion is a result of the scalings of the following quantities in the thermodynamic limit: $\rho\sim \phi \sim p\sim N^0, x\sim t\sim N$.

\bibitem{quan2007} H. T. Quan, Y. X. Liu, C. P. Sun, and F. Nori, Phys. Rev. E \textbf{76}, 031105 (2007).

\bibitem{Jaynes1957} E. T. Jaynes,  \emph{Information theory and statistical mechanics, I and II}. Phys. Rev. 1957, 106, 620.
\bibitem{Jaynes1983} \emph{E. T. Jaynes: Papers on Probability, Statistics and Statistical Physics}, edited by R. D.  Rosenkrantz,  (Reidel, Dordrecht,   1983).


\bibitem{Kaniadakis1996} G. Kaniadakis, A. Lavagno, P. Quarati, Nucl. Phys. B \textbf{466},  527 (1996).



\end{thebibliography}
\end{document}